\documentclass[final,5p,times,twocolumn,preprint]{elsarticle}
\usepackage{eucal} 
\usepackage{bm}
\usepackage{graphicx}
\usepackage[hypertexnames=false]{hyperref}
\usepackage{amsmath,amssymb,amsfonts,latexsym}
\allowdisplaybreaks
\newcommand{\fstate}{\rm f}
\newcommand{\istate}{\rm i}
\newcommand{\mstate}{\rm n}
\newcommand{\fistate}{\fstate\hspace{.04em}\istate}
\newcommand{\fmstate}{\fstate\hspace{.04em}\mstate}
\newcommand{\mistate}{\mstate\hspace{.04em}\istate}
\newcommand{\Sthreeistate}{\mathit{S}_{3\istate}}
\newcommand{\Sthreeistateprime}{{\mathit{S}_{3\istate}}^{\hspace{-.5em} \prime \hspace{.2em}}}
\newcommand{\Sthreeistateprimesub}{{\mathit{S}_{3\istate}}^{\hspace{-.4em} \prime \hspace{.2em}}}
\newcommand{\Bistateprime}{{B_{\istate}}^{\hspace{-.2em} \prime \hspace{.1em}}}
\begin{document}
\begin{frontmatter}
\title{Target normal single-spin asymmetry in inclusive electron-nucleon scattering \\
with two-photon exchange: Analysis using $1/N_c$ expansion}
\author[h,j]{J.~L.~Goity}
\ead{goity@jlab.org}
\author[j]{C.~Weiss}
\ead{weiss@jlab.org}
\author[m]{C.~T.~Willemyns}
\ead{cintiawillemyns@gmail.com}
\address[h]{Department of Physics, Hampton University, Hampton, VA 23668, USA}
\address[j]{Theory Center, Jefferson Lab, Newport News, VA 23606, USA}
\address[m]{Service de Physique Nucl\'eaire et Subnucl\'eaire,
Universit\'e de Mons, UMONS Research Institute for Complex Systems,
Place du Parc 20, 7000 Mons, Belgium}
\begin{abstract}
We calculate the target normal single-spin asymmetry caused by two-photon exchange in
inclusive electron-nucleon scattering in the resonance region. Our analysis uses the
$1/N_c$ expansion of low-energy QCD and combines $N$ and
$\Delta$ intermediate and final states using the contracted $\textit{SU}(4)$ spin-flavor symmetry.
The normal spin asymmetry obtained in leading-order accuracy in $1/N_c$
has magnitude $\sim$$10^{-2}$ and different sign in $ep$ and $en$ scattering.
It can be measured in electron scattering at lab energies $\sim$0.5--1.5 GeV
and provides a clean probe of two-photon exchange dynamics.
\end{abstract}
\end{frontmatter}
\section{Introduction}
Electron scattering represents a principal tool for exploring hadron structure and
strong interaction dynamics. The process is traditionally described in leading order of the
electromagnetic coupling (one-photon exchange approximation), where the amplitude is
proportional to the transition matrix element of the electromagnetic current operator
between the hadronic states. Recent developments in experiment and theory point to the
need of including higher-order interactions between the electron and the hadronic system
(two-photon exchange) in certain observables \cite{Carlson:2007sp}.
Measurements of the proton form factor ratio $G_E^p/G_M^p$ at Jefferson Lab
using the Rosenbluth separation and polarization transfer methods show discrepancies
that have been associated with two-photon
exchange \cite{JeffersonLabHallA:1999epl,Guichon:2003qm,Blunden:2003sp}.
A direct demonstration of two-photon exchange becomes possible through comparison
of electron and positron scattering in experiments at DESY
\cite{OLYMPUS:2016gso,OLYMPUS:2020dgl} and Jefferson Lab \cite{Accardi:2020swt}.
Two-photon exchange is also discussed in connection with muon scattering
at MUSE \cite{Cline:2021vlw} and plays an important role in radiative corrections
to parity-violating electron scattering \cite{Afanasev:2005ex}.
Two-photon exchange has thus become as field of research in its own right.

A particularly interesting observable is the target spin dependence in
inclusive electron-nucleon scattering,
\begin{align}
e(k_{\istate}) + \text{$\mathit{N}$$\uparrow$} (p_{\istate}, a_{\istate})
\rightarrow e(k_{\fstate}) + X(p_{\fstate}),
\label{process}
\end{align}
where $X$ denotes the hadronic final states accessible at the incident energy,
which are summed over. If the electron is unpolarized, and the nucleon is polarized
with a spin 4-vector $a_{\istate}$, with $a_{\istate}^2 = -1$ for complete polarization,
the dependence of the differential cross section on the nucleon spin is of the form
\cite{Afanasev:2007ii}
\begin{align}
\frac{d\sigma}{d\Gamma_{\fstate}} = \frac{d\sigma_U}{d\Gamma_{\fstate}}
- e_N^\mu a_{\istate \mu} \frac{d\sigma_N}{d\Gamma_{\fstate}} 
\label{dsigma}
\end{align}
($d\Gamma_{\fstate}$ denotes the invariant phase space element of the final electron and
will be specified below). Here $e_N$ is the normalized pseudovector formed from
the initial and final electron momenta and the initial nucleon momentum ($\epsilon^{0123} = 1$),
\begin{align}
e_N^\mu \equiv \frac{N^\mu}{\sqrt{-N^2}},
\hspace{.5em}
N^\mu \equiv -\epsilon^{\mu\alpha\beta\gamma} p_{\istate \alpha} k_{\fstate \beta}
k_{\istate \gamma},
\hspace{.5em}
e_N^2 = -1.
\label{e_N}
\end{align}
In the nucleon rest frame, $\bm{p}_{\istate} = 0$, the spin 4-vector is
$a_{\istate} = (0, 2\bm{S}_{\istate})$, with $|\bm{S}_{\istate}| = 1/2$ for complete
polarization, and $e_N$ is the normal vector to the scattering plane,
so that the cross section Eq.~(\ref{dsigma}) depends on the normal component of the
nucleon spin,
\begin{align}
e_N = (0, \bm{e}_N), \hspace{1em}
\bm{e}_N = \frac{\bm{k}_{\fstate}
\times \bm{k}_{\istate}}{|\bm{k}_{\fstate} \times \bm{k}_{\istate}|} , \hspace{1em} 
- e_N^\mu a_{\istate \mu} = 2 \bm{e}_N \cdot \bm{S}_{\istate}
\label{e_N_3d}
\end{align}
[this form applies in any frame where
the 3-momenta $\bm{k}_{\istate}, \bm{k}_{\fstate}$ and $\bm{p}_{\istate}$
lie in a plane, e.g.\ the electron-nucleon center-of-mass (CM) frame,
where $\bm{p}_{\istate} + \bm{k}_{\istate} = 0$].
The spin-dependent cross section in Eq.~(\ref{dsigma})
is zero in one-photon exchange approximation, as a consequence of
$P$ and $T$ invariance,
and represents a pure two-photon exchange observable \cite{Barut:1960zz,Christ:1966zz}.
It is proportional to the
imaginary (absorptive) part of the $eN \rightarrow eX$ two-photon exchange amplitude,
which is given by the product of on-shell matrix elements between the initial, intermediate,
and final electron-hadron states. Unlike the real (dispersive) part, the imaginary part of
the two-photon exchange amplitude is infrared--finite and can be considered separately
from real photon emission into the final state \cite{Afanasev:2007ii}.

Measurements of the normal spin asymmetry (the ratio of the $N$ and $U$ cross sections)
have been performed in deep-inelastic electron scattering (DIS) on proton
\cite{HERMES:2009hsi} and $^3$He targets \cite{Katich:2013atq}. Theoretical calculations
in this kinematics have employed the parton picture and QCD interactions and produced
a wide range of estimates \cite{Afanasev:2007ii,Metz:2006pe,Metz:2012ui,Schlegel:2012ve}.
Further measurements at few-GeV energies are planned at Jefferson Lab \cite{Grauvogel:2021btg}.
Calculations in the resonance region need to account for the contributions of individual
hadronic channels to the inclusive cross section, including elastic scattering and
resonance excitation, and require appropriate methods.

In this work we analyze the normal spin dependence of inclusive $eN$ scattering in the
resonance region using the $1/N_c$ expansion. The method organizes low-energy
dynamics (hadron masses, couplings, form factors) based on the scaling properties
in the limit of a large number of colors in QCD and has been successfully applied
in many areas of hadronic physics \cite{tHooft:1973alw,Witten:1979kh,Gervais:1983wq,Gervais:1984rc,Dashen:1993as,Dashen:1993jt,Dashen:1994qi}. Low-lying baryon states are organized
in multiplets of the emerging contracted $\textit{SU}(4)$ spin-flavor symmetry,
with the baryon masses
$O(N_c)$ and the splitting inside multiplets $O(N_c^{-1})$. The ground-state multiplet
contains the $N$ and $\Delta$, and transitions between them are governed by the
symmetry and can be computed using group-theoretical techniques,
with the parameters for the $N$--$\Delta$ and $\Delta$--$\Delta$ transitions
fixed in terms of the measurable $N$--$N$ transitions.

The $1/N_c$ expansion offers specific advantages for studying two-photon exchange
in inclusive scattering. The method treats $N$ and $\Delta$
states on the same basis and enables a consistent description of inelastic channels
and inclusive scattering in the resonance region. The group-theoretical techniques permit
efficient calculation of the sums over channels in intermediate and final states.
The parametric ordering of the kinematic variables gives rise to a physical picture
that enables an intuitive understanding of the two-photon exchange process.

In this letter we present the leading-order $1/N_c$ expansion and describe the
calculational techniques and physical picture specific to this situation. A full analysis,
including $1/N_c$ corrections and suppressed structures, will be presented in a
forthcoming article.
\section{Method}
\subsection{Kinematics and final states}
\label{subsec:kinematics}
Inclusive electron scattering Eq.~(\ref{process}) is characterized by three
independent kinematic variables, corresponding to the incident energy, the momentum
transfer, and the energy transfer of the process. They can be chosen as the invariant variables
\begin{align}
s &\equiv (k_{\istate} + p_{\istate})^2 = (k_{\fstate} + p_{\fstate})^2 ,
\label{s_def}
\\
t &\equiv (k_{\istate} - k_{\fstate})^2 = (p_{\fstate} - p_{\istate})^2 = q^2,
\label{t_def}
\\
m_X^2 &= (q + p_{\istate})^2 = p_{\fstate}^2,
\end{align}
where $q \equiv k_{\istate} - k_{\fstate} = p_{\fstate} - p_{\istate}$ is the 4-momentum transfer.
In the following we use the CM frame, where the 3-momenta in the initial and final states
are $\bm{p}_{\istate} = -\bm{k}_{\istate}, \bm{p}_{\fstate} = -\bm{k}_{\fstate}$, with
\begin{align}
|\bm{k}_{\istate}| &= \frac{s - m^2}{2\sqrt{s}}, \hspace{2em}
|\bm{k}_{\fstate}| = \frac{s - m_X^2}{2\sqrt{s}},
\label{pcm_from_s}
\\[1ex]
t &= -2 |\bm{k}_{\fstate}| |\bm{k}_{\istate}| (1 - \cos \theta ),
\hspace{2em} \theta \equiv \textrm{angle} (\bm{k}_{\fstate}, \bm{k}_{\istate}) ,
\label{t_cm}
\end{align}
where $m$ is the nucleon mass.

When analyzing the process Eq.~(\ref{process}) in the $1/N_c$-expansion, we have to specify
the scaling behavior of the kinematic variables in the parameter $1/N_c$. Different choices
are possible, leading to different types of expansions. Here we consider the domain where
the initial and final CM momenta are
\begin{align}
|\bm{k}_{\istate}|, |\bm{k}_{\fstate}| \; = \; O(N_c^0),
\label{domain_pcm}
\end{align}
corresponding to
$\sqrt{s} = O(N_c)$ and $\sqrt{s} - m = O(N_c^0)$.
For the final-state masses we consider values such that
\begin{align}
m_X - m \; = \; O(N_c^{-1}),
\hspace{2em}
m, m_X \; = \; O(N_c).
\label{domain_mx}
\end{align}
In this domain the only accessible final states are the ground-state baryon multiplet
containing the $N$ and $\Delta$ states, $X = N + \Delta$;
other baryon multiplets have masses $m_X - m = O(N_c^0)$
and are not accessible as final states. Together, Eqs.~(\ref{domain_pcm})
and (\ref{domain_mx}) imply
\begin{align}
|\bm{k}_{\istate}| - |\bm{k}_{\fstate}| \; = \; \frac{m_X^2 - m^2}{2\sqrt{s}}
\; = \; O(N_c^{-1})  \; \ll \; |\bm{k}_{\fstate}|, |\bm{k}_{\istate}| .
\label{domain_pcm_difference}
\end{align}
In leading order of $1/N_c$ we can therefore neglect the difference
between $|\bm{k}_{\istate}|$ and $|\bm{k}_{\fstate}|$ and use the common CM momentum
\begin{align}
k \; \equiv \; |\bm{k}_{\istate}| \; = \; |\bm{k}_{\fstate}| + O(N_c^{-1}).
\label{k_def}
\end{align}
For the CM scattering angle we consider values
$\theta = O(N_c^0)$, which together with Eq.~(\ref{domain_pcm}) implies
\begin{align}
t = O(N_c^0).
\end{align}

The parametric ordering in $1/N_c$ adopted here gives rise to a definite physical picture of
the scattering process. The electron with energy $O(N_c^0)$ scatters from the heavy nucleon
with mass $O(N_c)$, losing a small fraction $O(N_c^{-1})$ of its energy. The nucleon
remains intact or gets excited to a $\Delta$ by absorbing a small energy $O(N_c^{-1})$.
The velocity of the initial/final baryons is small $O(N_c^{-1})$, and their kinetic energy
is negligible compared to the electron energy. However, the momentum transfer is $O(N_c^0)$,
so that the process probes the internal structure of the baryons.

In the parametric domain considered here, inelastic scattering consists simply in the transition
from $N$ to $\Delta$ states, which can be regarded as stable in leading order of $1/N_c$
(the $\Delta$ width is suppressed). This corresponds to the physical situation that $\pi N$
final states are predominantly produced through $\Delta$ resonance decay. Non-resonant $\pi N$
states do not appear explicitly at leading order in $1/N_c$ in the domain considered here.
\subsection{Currents and amplitudes}
In the group-theoretical formulation of large-$N_c$ QCD, the $N$ and $\Delta$ are
states in the $\textit{SU}(4)$ multiplet of ground-state baryons, characterized by the spin/isospin
$S = I = 1/2$ and $3/2$, the spin projection $S_3$, and the
isospin projection $I_3$, denoted collectively by $B \equiv \{S, S_3, I_3 \}$.
The electron scattering process takes the form of a transition between baryon states
$\langle B_{\fstate} | ... | B_{\istate} \rangle$.
We denote the electron-baryon scattering amplitude by
\begin{align}
M(k, \bm{n}_{\fstate}, \bm{n}_{\istate} | \lambda, B_{\fstate}, B_{\istate})
\; \equiv \; M_{\fistate} ,
\label{amplitude}
\end{align}
where $k$ is the common CM momentum Eq.~(\ref{k_def}), and
\begin{align}
\bm{n}_{\istate} \equiv \bm{k}_{\istate}/ |\bm{k}_{\istate}|,
\hspace{2em}
\bm{n}_{\fstate} \equiv \bm{k}_{\fstate}/ |\bm{k}_{\fstate}|
\end{align}
are the unit vectors along the initial/final electron CM momenta.
In our convention the electron states have covariant normalization, while the baryon states have
non-covariant normalization; in this way the baryon mass does not appear in the
expressions for the phase space integral Eq.~(\ref{amplitude_e4}) and cross section
Eq.~(\ref{cross_section_general}), which is natural for the $1/N_c$ expansion.
In Eq.~(\ref{amplitude}), $\lambda$ is the electron helicity (spin projection
on $\bm{n}_{\istate}$ and $\bm{n}_{\fstate}$), which is conserved in the scattering process.
The baryon spins are quantized along a fixed direction in the CM frame; in this way the
initial and final states have the same quantization axis, and the spin transitions can be
computed using algebraic identities \cite{Dashen:1993jt,Dashen:1994qi}.\footnote{The following
calculation does not refer to a specific coordinate system. For definiteness we can imagine
using a system where $\bm{n}_{\fstate} + \bm{n}_{\istate}$ defines the $+x$-direction,
and $\bm{n}_{\fstate} - \bm{n}_{\istate}$ the $+z$-direction, and quantize the baryon spin
along the $+z$-direction; in this system the normal vector $\bm{e}_N$,
Eq.~(\ref{e_N_3d}), points in the $+y$ direction, and the spin density matrix
Eq.~(\ref{rho_N}) is $\sigma^y/2$.}

The amplitude Eq.~(\ref{amplitude}) can be computed as an
expansion in the electromagnetic coupling,
\begin{align}
M_{\fistate} = M_{\fistate}^{(e2)} + M_{\fistate}^{(e4)} + \ldots
\label{amplitude_expanded}
\end{align}
The $e^2$ term (one-photon exchange) is given by the product of the electron and
baryon currents,
\begin{align}
M_{\fistate}^{(e2)} &= -\frac{e^2}{t_{\fistate}} (j^\mu)_{\fistate} (J_{\mu})_{\fistate} ,
\label{amplitude_e2}
\\
t_{\fistate} &= -2 k^2 (1 - \bm{n}_{\fstate}\bm{n}_{\istate}) ,
\\[1ex]
(j^\mu)_{\fistate}
&=
\bar u (\bm{n}_{\fstate}, \lambda) \, \gamma^\mu \, u(\bm{n}_{\istate}, \lambda),
\label{current_electron}
\\[1ex]
(J^\mu)_{\fistate} &= \langle {-\bm{n}_{\fstate}}, B_{\fstate}| \hat{J}^\mu |
{-\bm{n}_{\istate}}, B_{\istate} \rangle .
\label{current_baryon_general}
\end{align}
The minus sign in Eq.~(\ref{amplitude_e2}) comes from the negative electric charge of the
electron. The electron current Eq.~(\ref{current_electron}) is the standard current of
the spin-1/2 particle; its explicit form can be derived from the spinors in the CM frame.
The baryon current Eq.~(\ref{current_baryon_general}) can be constructed using the
large-$N_c$ $\textit{SU}(4)$ spin-flavor symmetry and expanded in the generators
$\{ \hat{1}, \hat{I}^a, \hat{S}^i, \hat{G}^{ia}\} (i, a = 1,2,3)$
\cite{Dashen:1993jt,Dashen:1994qi,Fernando:2019upo}. Their matrix elements are of the order
\begin{align}
\langle B_{\fstate}| \{ \hat{1}, \hat{I}^a, \hat{S}^i \}  |B_{\istate}\rangle = O(N_c^0),
\hspace{2em} \langle B_{\fstate}| \hat{G}^{ia}  |B_{\istate}\rangle = O(N_c).
\end{align}
The full $1/N_c$ expansion of the current is given in Ref.~\cite{Fernando:2019upo}.
In the present study
we focus on the leading-order contribution to the cross sections, which is produced by
the isovector magnetic current proportional to $\hat{G}^{i3}$. This current is given by
\begin{align}
(J^0)_{\fistate} = 0,
\hspace{1em}
(J^i)_{\fistate} = 
i k \,
\epsilon^{ijk} \, (n_{\fstate} - n_{\istate})^j \,
\langle B_{\fstate}| \hat{G}^{k3} |B_{\istate}\rangle \, F(t_{\fistate}) .
\label{current_G}
\end{align}
It satisfies the transversality condition $q^\mu (J_\mu)_{\fistate} = 0$ for all
transitions between multiplet states, without corrections in $1/N_c$.

The function $F(t)$ in Eq.~(\ref{current_G}) (dimension $\textrm{mass}^{-1}$) 
is the large-$N_c$ form factor, which describes the dynamical response of the large-$N_c$
baryon to the momentum transfer $t = O(N_c^0)$. It can be determined by matching
the $N \rightarrow N$ matrix element of the large-$N_c$ current Eq.~(\ref{current_G})
with the physical nucleon current at $N_c = 3$. At leading order in $1/N_c$
one obtains
\begin{align}
F(t) \; = \; \left. \frac{G_M^V(t)}{m} \right|_{\rm phys} ,
\label{matching}
\end{align}
where $G_M^V(t) \equiv \frac{1}{2}[G_M^p(t) - G_M^n(t)]$ is the physical nucleon
isovector magnetic form factor, whose value at $t=0$ is given by the proton and neutron
magnetic moments, $G_M^V(0) = \frac{1}{2} (\mu^p - \mu^n)$, and $m$ is
the physical nucleon mass. In this way the spin-flavor symmetry
fixes the $N$--$\Delta$ and $\Delta$--$\Delta$ form factors
in terms of the empirical $N$--$N$ form factor, showing the predictive
power of the $1/N_c$ expansion.
%
%
\begin{figure}[t]
\hspace{.1\columnwidth}
\includegraphics[width=.7\columnwidth]{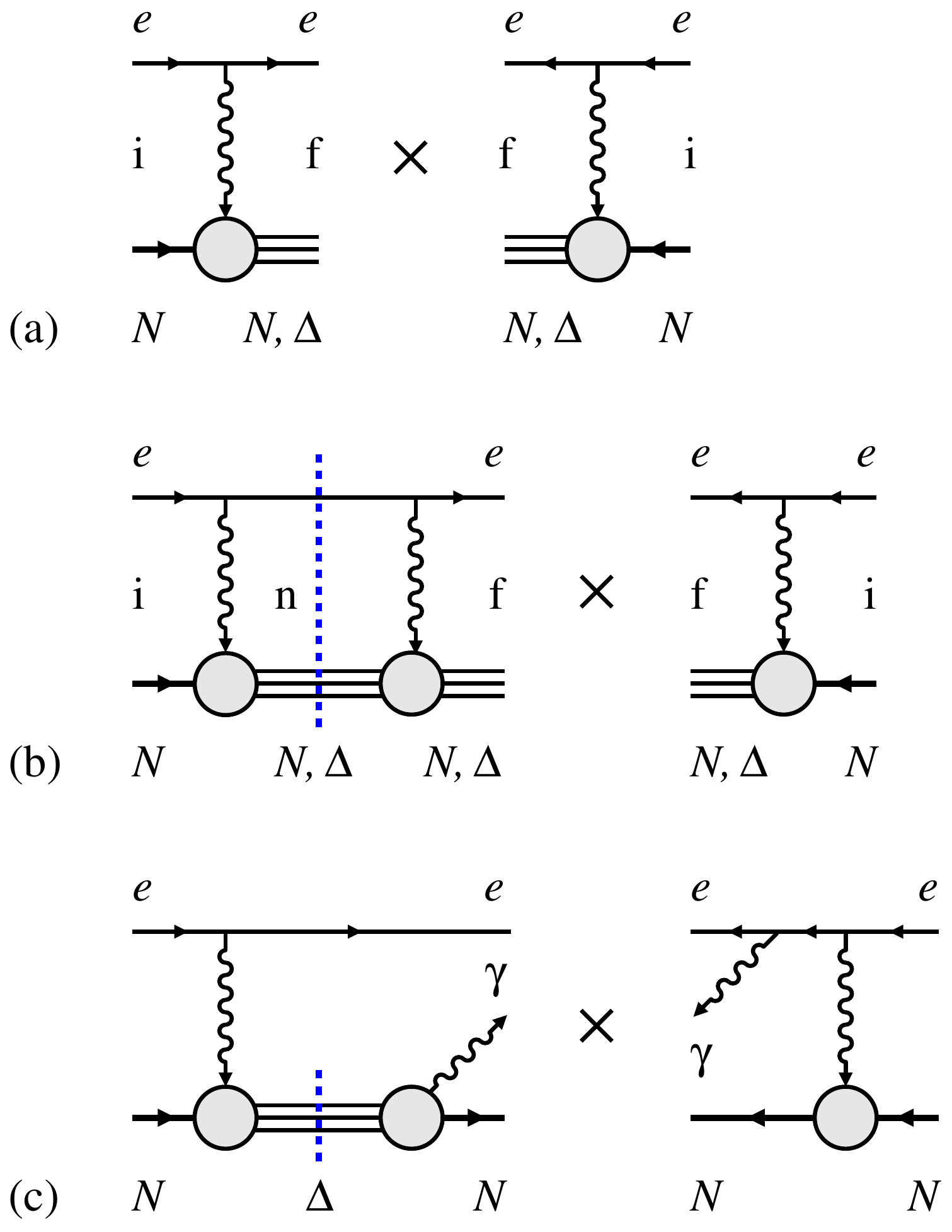}
\caption[]{Inclusive $eN$ scattering in the $1/N_c$ expansion in the domain
Eqs.~(\ref{domain_pcm}) and (\ref{domain_mx}).
(a)~Spin-independent cross section from square of $e^2$ amplitudes.
(b)~Spin-dependent cross section from interference of $e^4$ and $e^2$ amplitudes.
(c)~Interference of real photon emission from electron and baryon.}
\label{fig:diagrams}
\end{figure}

The $e^4$ term in the electron-baryon scattering amplitude Eq.~(\ref{amplitude_expanded})
results from two-photon exchange interactions. The absorptive part arises
from on-shell rescattering and can be computed as the product of two $e^2$ amplitudes,
integrated over the phase space of the intermediate state (see Fig.~\ref{fig:diagrams}b),
\begin{align}
M_{\fistate}^{(e4)} &= \frac{ik}{4\pi} \;
\int \frac{d\Omega_{\mstate}}{4\pi}
\; \sum_{B_{\mstate}} M_{\fmstate}^{(e2)} M_{\mistate}^{(e2)} .
\label{amplitude_e4}
\end{align}
We use the shorthand notation Eq.~(\ref{amplitude}) for the amplitudes of the
$\istate \rightarrow \mstate$ and $\mstate \rightarrow \fstate$ transitions. The integral is over
the momentum direction $\bm{n}_{\mstate}$ in the intermediate state, and the summation
is over the full set of baryon quantum numbers $B_{\mstate}$,
including $N$ and $\Delta$ states.
The prefactor in Eq.~(\ref{amplitude_e4}) is specific to our convention for the
amplitude (see above).

Some comments are in order regarding the inelasticity in the intermediate states
of the two-photon exchange amplitude. In the parametric domain
considered here, the scattering energy is $\sqrt{s} - m = O(N_c^0)$,
so that the intermediate states in principle include baryons with masses
$m_B - m = O(N_c^0)$ ($N^\ast$ states), larger than those of the final states
with $m_X - m = O(N_c^{-1})$.
However, the electromagnetic couplings of such $N^\ast$ states
to the ground state multiplet are suppressed by $1/\!\sqrt{N_c}$ relative to those
between ground state baryons \cite{Goity:2004pw,Goity:2005gs}.
In leading order of the $1/N_c$ expansion it is thus justified to retain only ground
state baryons $N$ and $\Delta$ as intermediate states.
Note also that the two-photon exchange amplitude Eq.~(\ref{amplitude_e4})
is free of collinear divergences, because the large-$N_c$ baryon currents in the
$\istate \rightarrow \mstate$ and $\mstate \rightarrow \fstate$
amplitudes satisfy the transversality conditions without corrections in
$1/N_c$ \cite{Afanasev:2007ii}.
\subsection{Cross section}
The cross section for inclusive $eN$ scattering Eq.~(\ref{process}) in the
$1/N_c$ expansion in the domain Eq.~(\ref{domain_pcm}) and (\ref{domain_mx})
is obtained from the amplitude Eq.~(\ref{amplitude}) as
\begin{align}
\frac{d\sigma}{d\Omega_{\fstate}} &= \frac{1}{16 \pi^2} \;\frac{1}{2} \sum_{\lambda}
\; \sum_{\Sthreeistateprimesub \Sthreeistate} \rho (\Sthreeistate, \Sthreeistateprime )
\nonumber
\\
& \times \sum_{B_{\fstate}}
M_{\fistate}^*(\lambda, B_{\fstate}, \Bistateprime) \,
M_{\fistate}(\lambda, B_{\fstate}, B_{\istate}) .
\label{cross_section_general}
\end{align}
We write the cross section as differential in the solid angle of $\bm{n}_{\fstate}$,
similar to elastic scattering. The inclusive scattering is expressed through the
summation over the final baryon states $B_{\fstate} = N, \Delta$. The initial baryon
is a nucleon,
$B_{\istate} = \{ {\textstyle\frac{1}{2}}, \Sthreeistate, I_{3\istate} \}$
and
$\Bistateprime = \{ {\textstyle\frac{1}{2}}, \Sthreeistateprime, I_{3\istate} \}$,
with $I_{3\istate} = \pm \frac{1}{2}$ for proton/neutron. The spin projections
are averaged with the nucleon spin density matrix $\rho$,
normalized as $\textrm{tr} \, \rho = 1$, which consists of an unpolarized
and a polarized part, $\rho = \rho_U + \rho_N$. The unpolarized part is
\begin{align}
\rho_U &= \frac{1}{2}\delta (\Sthreeistate, \Sthreeistateprime ) .
\end{align}
In the case of polarization along the unit vector $\bm{e}_N$, Eq.~(\ref{e_N}),
the polarized part is ($\sigma^k$ are the Pauli matrices)
\begin{align}
& \rho_N = \frac{1}{2} e_N^k \sigma^k (\Sthreeistate, \Sthreeistateprime ) ,
\label{rho_N}
\end{align}
such that the expectation value of the spin operator is 
\begin{align}
& \sum_{\Sthreeistateprimesub \Sthreeistate} \rho_N(\Sthreeistate, \Sthreeistateprime ) \;
\langle \Sthreeistateprime | \, \hat{S}^k \, | \Sthreeistate \rangle 
= \frac{1}{2} e_N^k .
\label{density_average_spin}
\end{align}

The spin-independent cross section of Eq.~(\ref{dsigma}) is obtained from the product of
$e^2$ amplitudes in Eq.~(\ref{cross_section_general}) (see Fig.~\ref{fig:diagrams}a),
\begin{align}
\frac{d\sigma_U}{d\Omega_{\fstate}}
\; &= \; \frac{1}{16 \pi^2} \; \frac{1}{2} \sum_{\lambda} \,
\sum_{\Sthreeistateprimesub \Sthreeistate} \rho_U
\sum_{B_{\fstate}} \,
M_{\fistate}^{(e2)\ast} M_{\fistate}^{(e2)};
\label{unpolarized_result_general}
\end{align}
the expression will be evaluated further below. For the spin-dependent cross section,
one can easily verify that it is zero at the same order in $e^2$,
because the $e^2$ amplitude is real and the average with
Eq.~(\ref{rho_N}) requires an imaginary part in one of
the amplitudes \cite{Barut:1960zz,Christ:1966zz}. The spin-dependent cross section appears instead
from the product of $e^2$ and $e^4$ amplitudes, i.e., the interference of
one- and two-photon exchange (see Fig.~\ref{fig:diagrams}b)
\begin{align}
\frac{d\sigma_N}{d\Omega_{\fstate}}
\; &= \;
\frac{1}{16 \pi^2} \;
\frac{1}{2} \sum_{\lambda} \;
\sum_{\Sthreeistateprimesub \Sthreeistate} \rho_N
\nonumber \\
& \times
\sum_{B_{\fstate}} \left[ M_{\fistate}^{(e2)\ast} M_{\fistate}^{(e4)} + M_{\fistate}^{(e4)\ast} M_{\fistate}^{(e2)} 
\right] .
\label{polarized_result_general}
\end{align}
With the $e^4$ amplitude given by Eq.~(\ref{amplitude_e4}), the spin-dependent cross section
is completely expressed in terms of the $e^2$ amplitude Eq.~(\ref{amplitude_e2}), and thus
in terms of the large-$N_c$ baryon current matrix elements.
\section{Results}
\label{sec:results}
\subsection{Spin-dependent cross section and asymmetry}
We now extract the leading $1/N_c$ term of the spin-dependent cross section.
It results from the isovector magnetic current Eq.~(\ref{current_G}) proportional
to the spin-flavor generator $\hat{G}^{i3}$. The $e^2$ amplitude Eq.~(\ref{amplitude_e2})
produced by this current is
\begin{align}
M^{(e2)}_{\fistate} &=
\frac{e^2 F(t_{\fistate})}{1 - \bm{n}_{\fstate}\bm{n}_{\istate}}
\, b^i_{\fistate} \langle B_{\fstate} | \hat{G}^{i3} | B_{\istate} \rangle ,
\\[1ex]
b^i_{\fistate} 
&\equiv i \epsilon^{ijk} (n_{\fstate} - n_{\istate})^j \frac{j^k_{\fistate}}{2k} ,
\label{b_def}
\end{align}
where $j^k_{\fistate}$ is the spatial part of the electron current Eq.~(\ref{current_electron}).
The product of $e^2$ and $e^4$ amplitudes in Eq.~(\ref{polarized_result_general}) then
becomes
\begin{align}
M_{\fistate}^{(e2) *} M_{\fistate}^{(e4)}
&= \frac{i e^6 k}{4\pi} \! \int \! \frac{d\Omega_{\mstate}}{4\pi}
\frac{F_{\fistate} F_{\fmstate} F_{\mistate}
\, b^{k*}_{\fistate} \, b^j_{\fmstate} \, b^i_{\mistate}}
{(1 - \bm{n}_{\fstate}\bm{n}_{\istate})
(1 - \bm{n}_{\fstate}\bm{n}_{\mstate})
(1 - \bm{n}_{\mstate}\bm{n}_{\istate})}
\nonumber
\\[1ex]
& \times 
\sum_{B_{\fstate}} \sum_{B_{\mstate}}
\langle \Bistateprime | \hat{G}^{k3} | B_{\fstate} \rangle
\langle B_{\fstate} | \hat{G}^{j3} | B_{\mstate} \rangle \langle B_{\mstate} | \hat{G}^{i3} | B_{\istate} \rangle ,
\label{e2_e4_product_general}
\end{align}
where $F_{\fistate} \equiv F(t_{\fistate})$, etc.
It represents a sequence of isovector magnetic transitions, with a tensor structure
governed by the electron current and the transition geometry. We evaluate it
using algebraic methods based on $t$-channel angular momentum considerations.
For the intermediate states in the $e^4$ amplitude, we sum over $B_{\mstate} = N + \Delta$
using the completeness relation in the ground state representation,
\begin{align}
\sum_{B_{\mstate}}
| B_{\mstate} \rangle \langle B_{\mstate} | = 1 ,
\label{completeness}
\end{align}
and the product in the last line of Eq.~(\ref{e2_e4_product_general}) becomes
\begin{align}
\sum_{B_{\fstate}} \langle \Bistateprime | \hat{G}^{k3} | B_{\fstate} \rangle
\langle B_{\fstate} | \hat{G}^{j3} \hat{G}^{i3} | B_{\istate} \rangle .
\label{product_G}
\end{align}
For the final states, we distinguish two cases:

{\it (i) Nucleon final state, $B_{\fstate} = N$.} In this case the matrix element of
$\hat{G}^{i3} \hat{G}^{j3}$
in Eq.~(\ref{product_G}) is a $\frac{1}{2} \rightarrow \frac{1}{2}$ spin transition,
and the tensor formed by the operator product can only have $t$-channel angular
momentum $J = 0$ or 1. The $J = 1$ part is antisymmetric in $ij$ and suppressed in $1/N_c$,
because the commutator of the operators is $[\hat{G}^{i3}, \hat{G}^{j3}] = O(N_c^0)$.
The tensor can therefore be projected on $J = 0$, which in leading order in $1/N_c$ gives
\begin{align}
\hat{G}^{j3} \hat{G}^{i3} &\rightarrow
\frac{1}{3} \, \delta^{ji} \, \hat{G}^{l3} \hat{G}^{l3}
= \frac{1}{3} \, \delta^{ji} \, \frac{N_c^2}{16},
\end{align}
and Eq.~(\ref{product_G}) becomes
\begin{align}
\frac{1}{3} \delta^{ji} \, \frac{N_c^2}{16} \, \langle \Bistateprime | \hat{G}^{k3} | B_{\istate} \rangle .
\label{contraction_n}
\end{align}

{\it (ii) Sum of nucleon and Delta final states, $B_{\fstate} = N + \Delta$.} In this case the summation
over $B_{\fstate}$ can be performed with the completeness relation, see Eq.~(\ref{completeness}),
and Eq.~(\ref{product_G}) becomes
\begin{align}
\langle \Bistateprime | \hat{G}^{k3} \hat{G}^{j3} \hat{G}^{i3} | B_{\istate} \rangle \; \equiv \; T^{kji} .
\end{align}
Because the commutator of the $\hat{G}^{i3}$ operators is suppressed in $1/N_c$, see above,
the tensor $T^{kji}$ can be regarded as completely symmetric in leading order.
As such it can be projected on overall $J = 1$ using
\begin{align}
T^{kji} \; &\rightarrow \; \frac{1}{5} (\delta^{kj} T^i + \delta^{ki} T^j + \delta^{ji} T^k) ,
\\
T^k \; &\equiv \; T^{kll} \; = \; 
\frac{N_c^2}{16} \, 
\langle \Bistateprime | \hat{G}^{k3} | B_{\istate} \rangle .
\label{contraction_n_delta}
\end{align}

The two cases thus lead to similar contractions of the tensor Eq.~(\ref{product_G}).
The remaining matrix element of $\hat{G}^{k3}$ in Eqs.~(\ref{contraction_n})
and (\ref{contraction_n_delta}) is proportional to the initial nucleon spin and
isospin, and in leading order of $1/N_c$ evaluates to
\begin{align}
\langle \Bistateprime | \hat{G}^{k3} | B_{\istate}\rangle &= \frac{N_c}{6}
\langle \Sthreeistateprime | \hat{S}^k | \Sthreeistate \rangle
\, (2 I_{3\istate}),
\end{align}
which can be averaged with the spin density matrix using Eq.~(\ref{density_average_spin}).
Altogether, we obtain the spin-dependent cross section in leading order of $1/N_c$
\begin{align}
\frac{d\sigma_N}{d\Omega_{\fstate}}
=& 
\frac{(2 I_{3\istate}) \, \alpha^3 N_c^3 k \, F_{\fistate}}
{96 \, (1 - \bm{n}_{\fstate}\bm{n}_{\istate})} \! \int \! \frac{d\Omega_{\mstate}}{4\pi}
\frac{F_{\fmstate} F_{\mistate} \, \Phi}{
(1 - \bm{n}_{\fstate}\bm{n}_{\mstate})
(1 - \bm{n}_{\mstate}\bm{n}_{\istate})} ,
\label{polarized_result}
\end{align}
\vspace{-3ex}
\begin{align}
\Phi =& \; \textrm{Re}\, \frac{1}{2} \sum_{\lambda} 
\frac{i}{3}  e_N^k
b^{k*}_{\fistate} b^{l}_{\fmstate} b^{l}_{\mistate}
\hspace{1em} (\textrm{for $N$ final state}),
\label{Phi_N}
\\[1ex]
\Phi =& \; \textrm{Re}\, \frac{1}{2} \sum_{\lambda} 
\frac{i}{5} e_N^k
\left(
  b^{k*}_{\fistate} b^{l}_{\fmstate} b^{l}_{\mistate}
+ b^{l*}_{\fistate} b^{k}_{\fmstate} b^{l}_{\mistate}
+ b^{l*}_{\fistate} b^{l}_{\fmstate} b^{k}_{\mistate}
\right)
\nonumber  \\
& \; (\textrm{for $N+\Delta$ final state}).
\label{Phi_N_Delta}
\end{align}
Here $\alpha \equiv e^2/4\pi$ is the fine structure constant.
The angular functions $\Phi$ can be evaluated using the explicit form of
the axial vectors $b^i_{\fistate}$ etc., Eq.~(\ref{b_def}).
The spin-dependent cross section Eq.~(\ref{polarized_result})
is proportional to the initial nucleon isospin $(2 I_{3\istate}) = \pm 1$
and has different sign for $ep$ and $en$ scattering
\begin{align}
d\sigma_N = d\sigma_N [ep] =  -d\sigma_N [en].
\end{align}

We also compute the spin asymmetry
\begin{align}
A_N \; &\equiv \; \left. \frac{d\sigma_N}{d\Omega_{\fstate}}
\right/ \frac{d\sigma_U}{d\Omega_{\fstate}} ,
\label{asymmetry_results}
\end{align}
by dividing by the unpolarized cross section computed in the same approximation.
In leading order of $1/N_c$, the unpolarized cross section Eq.~(\ref{unpolarized_result_general})
arises from the isovector magnetic current in the $e^2$ amplitude. In the case of
summation over $N$ and $\Delta$ final states, $B_{\fstate} = N + \Delta$, the result is
\begin{align}
\frac{d\sigma_U}{d\Omega_{\fstate}}
\; &= \; 
\frac{\alpha^2 \, N_c^2 \, F_{\fistate}^2}
{24 \, (1 - \bm{n}_{\fstate}\bm{n}_{\istate})^2} \;
\frac{1}{2}\sum_{\lambda} b^{i\ast}_{\fistate} b^i_{\fistate}
\\[1ex]
&= \; 
\frac{\alpha^2 \, N_c^2 \, (3 - \bm{n}_{\fstate} \bm{n}_{\istate}) \, F_{\fistate}^2}
{48 \, (1 - \bm{n}_{\fstate}\bm{n}_{\istate})} .
\label{unpolarized_result}
\end{align}
The spin-independent cross section in this approximation is independent
of the initial nucleon isospin; the asymmetry Eq.~(\ref{asymmetry_results})
therefore has the same isospin dependence as the spin-dependent cross section
in the numerator,
\begin{align}
A_N = A_N [ep] =  -A_N [en].
\end{align}

Some comments on these result from the perspective of the $1/N_c$ expansion are in order.
First, the spin-dependent cross section is parametrically large in $N_c$, as it
arises from the maximal product of isovector magnetic currents with matrix
elements $O(N_c)$. Second, our calculation provides an example of the
``$I = J$ rule'' of large-$N_c$ QCD, according to which leading structures appear
with $t$-channel quantum numbers $I = J$ \cite{Mattis:1988hf,Mattis:1988hg,Lebed:2006us}.
The spin-dependent cross section,
as a matrix element between the initial nucleon states $\langle \Bistateprime |... | B_i\rangle$,
is a structure with overall $J = 1$, and its leading large-$N_c$ result has $I = 1$.
It arises as the product of an $e^2$
amplitude with $I = J = 1$ (for both $N$ and $\Delta$ final states) with an $e^4$ amplitude
that is either projected on $I = J = 0$ (for $N$ final) or on
$I = J = 2$ (for $\Delta$ final), as can be observed in the algebraic calculation above.

\subsection{Numerical results}
%
%
\begin{figure}[t]
\includegraphics[width=.95\columnwidth]{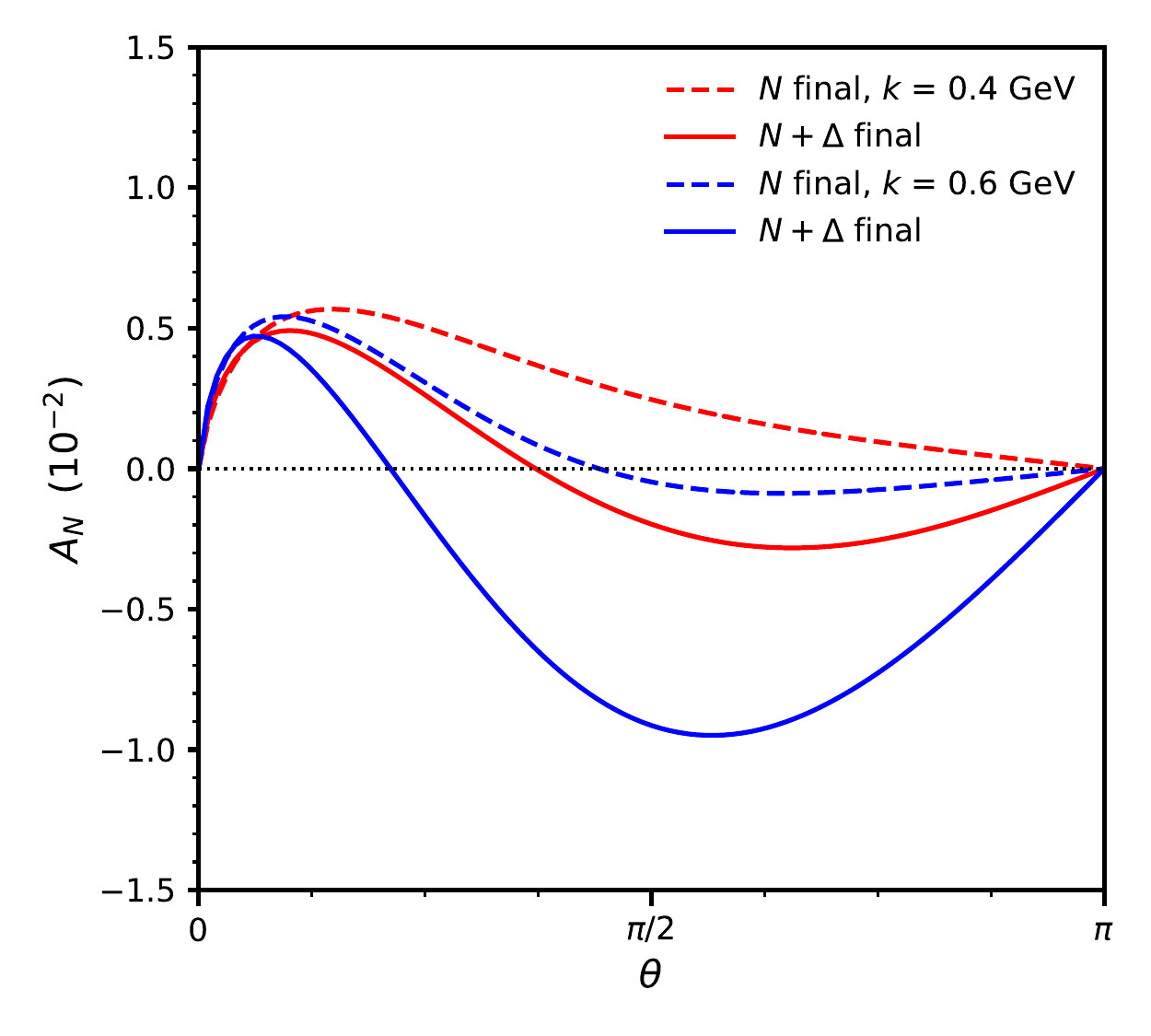}
\caption[]{Target normal single-spin asymmetry $A_N$ in inclusive $eN$ scattering, 
Eq.~(\ref{asymmetry_results}), in leading order of the $1/N_c$ expansion,
for two values of the CM momentum $k$, as a function of the CM scattering angle $\theta$.
\textit{Dashed lines:} $A_N$ with $N$ final state in $\sigma_N$ in the numerator.
\textit{Solid lines:} $A_N$ with $N + \Delta$ final states in $\sigma_N$.
In both cases, $\sigma_U$ in the denominator is with $N + \Delta$ final states.}
\label{fig:asymmetry}
\end{figure}
We now evaluate the asymmetry numerically and study its kinematic dependence using the
leading-order $1/N_c$ expansion results, Eqs.~(\ref{polarized_result})--(\ref{Phi_N_Delta})
and Eq.~(\ref{unpolarized_result}). The large-$N_c$ form factors $F(t)$
appearing in the expressions are fixed by the matching condition Eq.~(\ref{matching}),
and we use the standard dipole form $(1 - t/0.71 \, \textrm{GeV}^2)^{-2}$ to model
the empirical $t$-dependence.

Figure~\ref{fig:asymmetry} shows $A_N$ for two values of the CM momentum $k$,
as a function of the CM scattering angle $\theta = \textrm{angle}(\bm{n}_{\fstate}\bm{n}_{\istate})$.
Results are shown for the
cases of $N$ and $N + \Delta$ final states in $\sigma_N$ in the numerator;
$\sigma_U$ in the denominator is always for $N + \Delta$ final states;
in this way one can add/subtract the results for $A_N$ in the graph and
see the contributions of the various channels to $\sigma_N$. (The intermediate states
in the two-photon exchange amplitude in $\sigma_N$ are always the sum $N + \Delta$.)

One observes: (i)~$A_N$ vanishes at $\theta = 0$ and $\pi$, which is natural,
as at these angles the normal vector $\bm{n}_2 \times \bm{n}_1$ vanishes.
(ii)~The contribution of $\Delta$ final states (the difference of
the results for $N + \Delta$ and $N$ final states) is small at small $\theta$ but
becomes significant at $\theta \sim \pi/2$, causing the $A_N$ for final $N + \Delta$
to be several times larger than that for final $N$. (iii)~$A_N$ reaches values
$\sim 10^{-2}$ at $\theta \sim \pi/2$ and $k \sim 0.6$ GeV.

Some comments on the region of applicability of the large-$N_c$ expressions are in order.
(i)~The $1/N_c$ expansion in the domain Eq.~(\ref{domain_mx})
assumes that the $\Delta$ channel is open. The expressions should therefore
be applied at CM energies above the physical $\Delta$ threshold $\sqrt{s} = 1.23$ GeV.
(ii)~The calculation relies on the $1/N_c$ suppression of $N^\ast$ states with masses
$m_B - m = O(N_c^{0})$ as intermediate states in the two-photon exchange amplitude.
This should be reasonable at energies up to and moderately above the $N^\ast$
threshold $\sqrt{s} \sim$ 1.5 GeV, but not substantially above it. 
(iii)~The leading-order results for $\sigma_N$ and $A_N$ arise entirely
from magnetic currents, which are proportional to the momentum transfers
at the vertices. They are not expected to be accurate at small $\theta \ll \pi/2$
and $k \ll$ 1 GeV, where the momentum transfers are kinematically suppressed
and contributions from electric currents are important (those can be computed as
part of the $1/N_c$ corrections). Altogether, we expect the leading-order $1/N_c$
result to be a fair approximation at CM momenta $k \sim$ 0.3--0.6 GeV and
large angles $\theta \sim \pi/2$. In this kinematics the accuracy of the leading-order
$1/N_c$ result is naively estimated to be of the order $\sim 1/3$, as observed in other
hadronic observables. A more quantitative assessment of the accuracy will become possible
with the computation of $1/N_c$ corrections.

We note that the numerical results for $\sigma_N$ and $A_N$ in the $1/N_c$ expansion
are strongly affected by the presence of the form factors in the integral in
Eq.~(\ref{polarized_result}). This indicates that the two-photon
exchange observables are sensitive to baryon structure in the domain considered here.
\section{Extensions}
We have studied the target normal single-spin asymmetry in inclusive $eN$ scattering
in leading order of the $1/N_c$ expansion, in the parametric domain where the energy
transfer is $O(N_c^{-1})$ and allows for $\Delta$ excitation,
and the momentum transfer is $O(N_c^{0})$ and probes the internal structure of the baryons. 
The results can be extended and applied in several ways.

The method developed here, particularly the algebraic approach in Sec.~\ref{sec:results},
can be used to compute $1/N_c$ corrections to the leading-order result.
These corrections will quantify the numerical accuracy of the
leading-order result for the isovector $\sigma_N$, and provide estimates of the
isoscalar $\sigma_N$, which appears only at subleading order.

In the parametric domain considered here, the intermediate states in the two-photon exchange
amplitude have energies $\sqrt{s} - m = O(N_c^{0})$ and include $N^\ast$ baryons with masses
$m_B - m = O(N_c^{0})$. In the present analysis we have neglected such intermediate states
because their electromagnetic couplings to the initial/final $N$ and $\Delta$ states are
suppressed by factors $1/\!\sqrt{N_c}$. When such $N^\ast$ states are included, they can enhance the
region of quasi-real photon exchange (collinear to the electron momenta) in the two-photon
exchange integral, which could result in numerically enhanced contributions. This effect
needs to be analyzed in the context of the higher-order $1/N_c$ expansion.

The cross section for inclusive $eN$ scattering includes also real photon emission
into the final state (Fig.~\ref{fig:diagrams}c). This process can be analyzed in the
$1/N_c$ expansion in the same manner as two-photon exchange (Fig.~\ref{fig:diagrams}b).
If the intermediate state is a $\Delta$, the emitted photon momentum is $k_\gamma = O(N_c^{-1})$,
because its energy is given by the mass difference $m_\Delta - m = O(N_c^{-1})$.
In this situation the coupling through the leading magnetic vertex is suppressed,
and we expect real photon emission to be suppressed in leading order of the $1/N_c$ expansion.
If the intermediate state is a $N^\ast$ with mass difference $m_B - m = O(N_c^0)$,
as becomes possible in higher orders in $1/N_c$, the emitted photon momentum is
$k_\gamma = O(N_c^0)$, and real emission can contribute at the same order
as two-photon exchange. The higher-order $1/N_c$ expansion therefore needs to treat
two-photon exchange and real photon emission on the same basis.
Overall, the parametric expansion in $1/N_c$ provides definite prescriptions for
including both $N^\ast$ excitation and real photon emission in the inclusive normal
single-spin asymmetry.

The $1/N_c$ expansion can also be performed in parametric domains different from
Eqs.~(\ref{domain_pcm}) and (\ref{domain_mx}). For example, the choice
$k = O(N_c^{-1})$ leads to a ``low-energy expansion'' in which the
electric currents enter in the same order as the magnetic ones,
giving rise to a different physical picture.

The framework of the $1/N_c$ expansion can also be used to explore the transition between the
resonance and DIS regions and the realization of quark-hadron duality in two-photon exchange
observables. Theoretical estimates of $A_N$ differ by 1-2 orders of magnitude
between the resonance and DIS regions, because of the large effects of the anomalous magnetic
moment that are present in resonance production but disappear
in DIS \cite{Afanasev:2007ii,Metz:2006pe,Metz:2012ui,Schlegel:2012ve}.
Performing $1/N_c$ expansions in different kinematic domains would help explain
how the transition happens.

The methods developed here can be applied to the beam spin asymmetry in $eN$
scattering, a two-photon exchange effect proportional to the electron mass, which is
being studied as a background to parity-violating electron scattering
\cite{Afanasev:2004pu,Carlson:2017lys,Koshchii:2019mgv}.

This material is based upon work supported by the U.S.~Department of Energy, 
Office of Science, Office of Nuclear Physics under contract DE-AC05-06OR23177 (JLG, CWe);
by the National Science Foundation, Grant Number PHY 1913562 (JLG);
and by the Fonds de la Recherche Scientifique (FNRS) (Belgium), Grant Number 4.45.10.08 (CTWi).
%
%
%
\bibliography{ssa_ncexp}
\end{document}